\documentclass[twocolumn,aps,pra,groupedaddress]{revtex4}
\usepackage{epsfig,amssymb,amsmath}
\def\comment#1{}
\def\togli#1{}
\def\labell#1{\label{#1}}
\def\tr{ {\rm tr}~ }
\def\ds{ \Sigma  }

\def\N{ {\cal N} }
\def\Th{ {\rm Th} }
\begin{document}
%\fbox{\scriptsize Preliminary draft}
%\fbox{{\scriptsize Submitted draft.}}
%\fbox{{\scriptsize Preprint quant-ph/.}}
%Title of paper
\title{ The bosonic minimum output entropy conjecture and Lagrangian
  minimization} \author{S. Lloyd$^{1,2}$, V. Giovannetti$^{3}$, L.
  Maccone$^{1,4}$ N.J. Cerf$^{5}$, S. Guha$^{6}$, R.
  Garcia-Patron$^1$, S.  Mitter$^1$, S. Pirandola$^1$, M.B.
  Ruskai$^7$, J.H. Shapiro$^{1,8}$, H. Yuan$^1$}\affiliation{1.~W.M.
  Keck Center for Extreme Quantum Information Theory, MIT.
  2.~Department of Mechanical Engineering, MIT 3-160, Cambridge MA
  02139. 3.~Scuola Normale Superiore, Pisa.  4.~Institute for
  Scientific Interchange, Torino. 5.~Universit\'e Libre de Bruxelles,
  Centre for Quantum Information and Communication. 6.~BBN
  Technologies Corporation, Cambridge.  7.~Department of Mathematics,
  Tufts University. 8.~Research Laboratory of Electronics, MIT. }
%\date{\today}

\begin{abstract}
  We introduce a new form for the bosonic channel minimal output
  entropy conjecture, namely that among states with equal input
  entropy, the thermal states are the ones that have slightest
  increase in entropy when sent through a infinitesimal thermalizing
  channel. We then detail a strategy to prove the conjecture through
  variational techniques. This would lead to the calculation of the
  classical capacity of a communication channel subject to thermal
  noise. Our strategy detects input thermal ensembles as possible
  solutions for the optimal encoding of the channel, lending support
  to the conjecture.  However, it does not seem to be able to exclude
  the possibility that other input ensembles can attain the channel
  capacity.
\end{abstract}
\pacs{} 
\maketitle 

Bosonic communication channels employ bosons such as photons or
phonons to carry information~\cite{2,3,4,5,6,HW,EW,CEGH,EXT,WOLF1,REN,CANONICAL1,CAP}.   Establishing the
fundamental communications capacity of such channels is a primary
outstanding scientific question~\cite{BENSHOR}. It is also a question of considerable
social relevance: optical communication channels are pressing down to
the level where significant amounts of information are carried by
individual photons.  It would be useful to know whether we are nearing
the limit of communication capacity, or whether quantum effects such
as squeezing and entanglement might be used to enhance communication
capacity significantly.

Ever since Shannon established the capacity of the continuous
classical communication channel with Gaussian noise and loss more than
half a century ago~\cite{1}, researchers have sought to establish the
capacity of the corresponding quantum channel, the bosonic Gaussian channel
that transmits information using bosons such as photons or phonons~\cite{2,3,4,5,6,CAP,HW,EW,CEGH,EXT,WOLF1,REN,CANONICAL1}.  
The classical capacity~\cite{HWS} of the bosonic channel with thermal
noise and linear loss hinges on the bosonic minimum output entropy
conjecture, which states that the vacuum input gives the minimum
output von Neumann entropy for a channel with thermal noise~\cite{5,6,REN}.  If this
conjecture is true, then the channel attains its capacity for
coherent state inputs \cite{2,3,4,5,6,REN,CAP}.  Here we present an
infinitesimal version of this conjecture  and detail a variational
strategy for its proof, using a double Lagrangian constrained
minimization. 

Since the input that gives the minimum output entropy is a pure
state that lies on the boundary of the input space, variational
techniques such a Lagrangian methods might at first seem difficult
to apply.  We circumvent this problem by looking for the minimum
output entropy for states with a fixed, non-zero input entropy, and
then take the limit in which that input entropy to zero.
We find that thermal states are solutions to the Lagrange equations
and represent local minima to the output entropy. 
Unfortunately, unless the gradients of the minimization constraints
are linearly independent, one cannot conclude that the minima
recovered through the Lagrangian technique are unique~\cite{hadley}.
Thus, even though the results obtained below are consistent with the
conjecture (namely input coherent states are identified as minima for
the channel output entropy), we cannot conclude that they are the
global minima.  Namely, we cannot exclude that other input states can
give even lower output entropy, thus becoming the constituent of an
encoding that achieves the channel capacity. Nonetheless, the approach
detailed here constitutes a progress in answering the oldest question
of quantum information theory, by casting the problem in a framework
that is completely different than what previously analyzed.

\section{Conjectures}\label{sec1}

Consider a bosonic channel with additive Gaussian thermal noise, e.g.  the additive classical noise channel,  the thermal-noise channel, or the amplifier channel of Ref.~\cite{HW}.
%Establishing the ultimate capacity of bosonic channels with Gaussian
%noise and linear loss hinges on the following conjecture~\cite{5,6,7}.
%
%Without loss of generality 
%in the following  we will focus on the special case of  single mode 
%channels with additive Gaussian thermal noise. Specifically we will analyze the channel which  transforms any input state $\rho$ by randomly displacing it in the phase space
%according to the mapping
%\begin{eqnarray}\label{PRIMA}
% \rho \longrightarrow \Phi_{\Delta E} (\rho) = \int d^2  \mu\;  P_{\Delta E}(\mu)
%D(\mu) \rho D^\dagger(\mu)\;, 
%\end{eqnarray}
% where the integration is performed over
%the the complex plane, $D(\mu) = \exp[\mu a^\dag - \mu^* a]$ is the displacement operator, and
%$P_{\Delta E}(\mu)$ is a zero mean, Gaussian distribution of width
%$\Delta E$.  This maps is typically referred to as the
%additive classical noise channel~\cite{HW} and proving the minimum output entropy conjecture for it, 
%it will allows one to prove it also for the larger set of thermal 
%channels~\cite{5}.
Establishing its ultimate classical capacity hinges on the following conjecture~\cite{5,6,REN,CAP},
\newline

$\bullet${}{\it Conjecture (finite):}  The vacuum input state for the channel  gives 
the minimum output entropy.   
\newline 

\noindent That is, to get the smallest output entropy, do nothing.  If
this conjecture is false, then quantum `tricks' could enhance channel
capacity.  By contrast, if this conjecture is indeed true, then
references~\cite{5,6} show that Gaussian channels with noise and
linear loss attain their maximum capacity for coherent state inputs.
While highly plausible, the bosonic minimum output entropy conjecture
has resisted proof for some time now. Exploiting the fact that the above maps have a semigroup structure and thus admits a Lindblad generator (e.g. see Ref.~\cite{BANE}), 
 we now extend this conjecture
also to the infinitesimal case. Namely we investigate the
infinitesimal version of the additive Gaussian thermal noise channels, i.e. 
\newline 

$\bullet${}{\it Conjecture (infinitesimal):} Among input states with a
fixed entropy, thermal states give the minimum rate of output entropy
increase for the infinitesimal additive noise channel.
\newline

\noindent A proof of the infinitesimal version of the conjecture would
establish the truth also of the finite version. In fact, the rate of
entropy increase under thermalization is non-negative.  If the
infinitesimal conjecture were proved, one would conclude that for pure
state inputs the vacuum is the zero-temperature thermal state and
gives the minimum rate of increase in entropy under thermalization.
Then, thermalization would take the vacuum through a sequence of a
non-zero temperature thermal states of increasing entropy, each of
which would yield the minimum rate of entropy increase for its
particular entropy.  The path that starts at the vacuum and passes
through a sequence of thermal states therefore would give the minimum
entropy increase also for non-infinitesimal thermalization, proving
also the finite version of the conjecture.

Here we detail a possible strategy for a proof using a Lagrangian
technique, enforcing entropy, energy, and dynamical constraints using
Lagrange multipliers~\cite{hadley,8}.  Defining a suitable Lagrangian $\ds$
and setting its variation $\delta \ds = 0$ allows us to identify the
potential minima of the output entropy $S$.  Symmetries of thermal
noise imply that displaced thermal states will also give the minimum
output entropy: within a set of connected minima, we pick out the one
that has the smallest energy. Among the set of minima connected with a
thermal input, thermal states are the one with minimum output energy.
If one could exclude that other minima exist, the conjectures would be
proved: there is a single set of connected minima related to input
thermal states. Unfortunately, this turns out to be extremely
complicated since the gradients of the constrains in the constrained
minimization on the energy are not linearly independent. In this case,
the Lagrange procedure is not guaranteed to identify all the
minima~\cite{hadley}, and one would have to exclude the presence of
other minima (potentially with lower output entropy) through other
means.

\section{Thermalization and entropy increase}\label{sec2}

Without loss of generality in the following  we will focus on the special case of  a  single mode 
channel which  transforms any input state $\rho$ by randomly displacing it in the phase space
according to the mapping
\begin{eqnarray}\label{PRIMA}
 \rho \longrightarrow \Phi_{\Delta E} (\rho) = \int d^2  \mu\;  P_{\Delta E}(\mu)
D(\mu) \rho D^\dagger(\mu)\;, 
\end{eqnarray}
 where the integration is performed over
the the complex plane, $D(\mu) = \exp[\mu a^\dag - \mu^* a]$ is the displacement operator, and
$P_{\Delta E}(\mu) \equiv \exp[ - |\mu|^2/\Delta E] / (\pi \Delta E)$ is a zero mean, Gaussian distribution of width
$\Delta E$ (here  $a,a^\dagger$  are  the annihilation and
creation operators of the mode). This maps is typically referred to as the
additive classical noise channel~\cite{HW} and proving the minimum output entropy conjecture for it, 
it will allows one to prove it also for the larger set of thermal 
channels~\cite{5}.

%
%The additive classical noise channel~\cite{HW} for a single mode can be
%represented as the completely positive transformation which maps the
%input states $\rho$ according to the mapping
%\begin{eqnarray}\label{PRIMA}
% \rho \; \longrightarrow \; \int d^2  \mu\;  P_{\Delta E}(\mu)
%D(\mu) \rho D^\dagger(\mu)\;, 
%\end{eqnarray}
% where the integration is performed over
%the the complex plane, $D(\mu) = \exp[\mu a^\dag - \mu^* a]$ is the displacement operator, and
%$P_{\Delta E}(\mu)$ is a zero mean, Gaussian distribution of width
%$\Delta E$.  

As mentioned in the previous section, we will work with the infinitesimal version $\Th$ of the map $\Phi_{\Delta E}$. It is obtained form Eq.~(\ref{PRIMA}) in the limit of  $\Delta E\ll 1$ and
can be expressed as the mapping  
\begin{eqnarray} \label{therma}
\Th(\rho) = \rho + \gamma \; \Delta t \;  \N(\rho) + {O}(\Delta t^2)\;,
\end{eqnarray}
with $\N$ being the super-operator describing 
the infinitesimal thermalization of a bosonic mode according to the Lindblad equation, 
%with annihilation and
%creation operators $a,a^\dagger$ is described by the process
\begin{eqnarray}
{d \rho \over dt}& =& {\gamma \over 2} \big(
2 a \rho a^\dagger - a^\dagger a \rho - \rho a^\dagger a
+
2 a^\dagger \rho a - a a^\dagger \rho - \rho a a^\dagger \big)
\nonumber\\&\equiv& \gamma {\cal N}(\rho).
\labell{1}\;
\end{eqnarray}
Here $\gamma > 0$ governs the rate at which photons are
added and subtracted from the mode: over a brief time $\Delta t \ll1/\gamma$,
the average increase in energy for zero-mean input states
is $\Delta E = \gamma \Delta t$, independent of $\rho$ as can be easily verified from~(\ref{therma}), i.e.
%Define
%$\Th(\rho) = \rho + \gamma \Delta t \N(\rho)$: then we have
\begin{eqnarray}
\tr [\Th(\rho) a^\dagger a] =
\tr [\rho a^\dagger a] + \Delta E +{O}(\gamma^2 \Delta t^2)\;.
\end{eqnarray} 
%It is straightforward to verify that~\eqref{1} yields the
%infinitesimal version of the additive classical noise channel.
 
Consider now an input state $\rho$ wiht von Neumann entropy  $S(\rho) = - \tr \rho \ln \rho =S_0$. 
Its  entropy increase over time $\Delta t$ can be computed as 
\begin{eqnarray}
&& \Delta S  = - \tr \Th(\rho) \ln \Th(\rho) 
+ \tr \rho \ln \rho  \nonumber\\&&=
 - \gamma \Delta t~ \tr \N(\rho) \ln \rho + O(\gamma^2 \Delta t^2). 
\labell{2}\;
\end{eqnarray}

To minimize the output entropy for fixed input entropy $S_0$, we
introduce a new positive matrix variable $z$ and use the method of
Lagrange multipliers to set $z = \Th(\rho)$, as well as enforcing the
input entropy constraint.  The Lagrange multiplier technique tells us
that if our Lagrangian is continuous and has continuous derivatives in
the open interior of the set of density matrices, then the absolute
minimum of entropy increase either occurs at a stationary point of the
Lagrangian or on the boundary of the set.  (In infinite-dimensional
Hilbert spaces, the gradient of the entropy is everywhere
discontinuous.  Accordingly, to insure that our Lagrangian is
continuous and has continuous derivatives, we truncate our input
Hilbert space at some high photon number $N$, find the minimum output
entropy for finite $N$, and take the limit $N\rightarrow \infty$.
Technical details of this truncation can be found in the Appendix.)
We will see that inputs on the boundary maximize entropy increase.
Accordingly, the global minimum must be a stationary point of the
Lagrangian.

Because the thermalization process of Eq.~\eqref{1} is covariant under displacements
in the phase space, each local minimum is connected to a manifold of
local minima with the same output entropy by continuous displacements.
Within the set of states with this output entropy, we pick out the
output state with the lowest energy.  The only minimum energy
stationary points of that the Lagrangian procedure returns are thermal
states. So that, if we could exclude other local minima, we could
conclude that the thermal state gives the global minimum of entropy
production.

\section{Inputs on the boundary maximize entropy increase}\label{sec3}

First we show that inputs on the boundary give maxima of output
entropy.  The boundary consists of density matrices with at least one
zero eigenvalue, i.e. density matrices with a non-zero kernel,
$K(\rho)$.  Let $P_\perp$ be the projector onto the support of $\rho$,
i.e., onto the subspace orthogonal to $K$.  From the formula for
entropy increase, equation~\eqref{2}, we see if infinitesimal
thermalization takes any state from subspace perpendicular to the
kernel, $K^\perp$, into the kernel, then the rate of entropy increase
diverges.  That is, for states on the boundary, the rate of entropy
increase diverges unless $a P_\perp a^\dagger \subseteq P_\perp, \quad
a^\dagger P_\perp a \subseteq P_\perp.$ But in the input Hilbert
space, there is no nonzero proper subspace that can fulfill both these
requirements.  Accordingly, states on the boundary give divergent
rates of entropy increase.

\section{Stationary points of the Lagrangian}\label{sec4}

Since it does not occur on the boundary, all minima of entropy
increase must occur for stationary points of our Lagrangian.  
Define the Lagrangian
\begin{eqnarray}
 \ds  &=&- \tr z \ln z + \tr \rho \ln \rho    
 - \alpha(\tr \rho -1) -\eta(\tr z -1)
\nonumber\\&& - \lambda( - \tr \rho \ln \rho - S_0) 
 - \tr \Lambda ( z - \Th(\rho) ).
\labell{3}\;
\end{eqnarray}
Here, the $\alpha$ and $\eta$ Lagrange multipliers enforce
the normalization constraints, $\lambda$ enforces the input
entropy constraint, and the Hermitian matrix of
Lagrange multipliers $\Lambda$ enforces the dynamical constraint.
 
Now find the stationary points of $\ds$.
Introduce small variations $z\rightarrow z + \delta z$,
$\rho \rightarrow \rho + \delta \rho$, together with similar
variations in the Lagrange multipliers.  We obtain
\begin{eqnarray}
 \delta \ds &=& ~ 
\tr \delta z\big( -\ln z -1  - \eta  - \Lambda
\big)
\nonumber\\&&  
+ \tr \delta \rho \big( (\lambda + 1)(\ln \rho + 1)
- \alpha 
 + \Th(\Lambda) \big)  
\nonumber\\&&  
 - \delta\alpha(\tr \rho -1) - \delta\eta(\tr z -1)
\nonumber\\&&  
 - \delta\lambda( - \tr \rho \ln \rho - S_0) 
 - \tr \delta\Lambda ( z - \Th(\rho) ).
\labell{4}\;
\end{eqnarray}
Here we have used the fact  
that $\tr[\Th(\delta \rho) \Lambda]$ $ = \tr [\delta \rho \Th(\Lambda)]$;
this is easily verified using equation~\eqref{1}.
Stationary points of $\ds$ 
correspond to points such that $\delta \ds = 0$,
which, upon simplification, yields the conditions,
\begin{eqnarray}
&& -\ln z  -1  - \eta - \Lambda= 0
\nonumber\\&&
 (\lambda +1)(\ln \rho + 1)  - \alpha +  \Th(\Lambda) =0
 \nonumber\\&&    \tr \rho = 1,\quad \tr z =1
 \nonumber\\&&
  - \tr \rho \ln \rho = S_0 \qquad
  z = \Th(\rho) .\labell{5}\;
\end{eqnarray}
These equations hold for any cutoff $N$ for the
truncated Hilbert space. 

Equation~\eqref{5} has many solutions.  It is not hard to show that,
in the limit $N\rightarrow \infty$, thermal states and displaced
thermal states solve equation~\eqref{5} for suitable choices of the
Lagrange multipliers.  On the face of it, equation~\eqref{5} might
also have additional solutions, unconnected with thermal states by
displacements, that could give lower output entropy than thermal
inputs. One strategy to rule out the possibility of such alternative
manifolds of minima is to perform a second minimization (on the
energy) and verify that only the minimum corresponding to thermal
input survives. 

Consider a manifold of connected states that give the same local
minimum of output entropy.  Within this manifold, we find the states
that minimize energy as well (because energy is bounded below, such
states always exist).  We show that thermal inputs and thermal outputs
minimize output entropy, and minimize energy within the the set of
states with that same output entropy.  If one could conclude that
there is {\it only} one manifold of local minima (the one connected to
the input thermal state with the specified input entropy), then
thermal inputs would give the {\it unique} local minimum of output
entropy.  Unfortunately we have not been able to prove such unicity.

\section{Stationary points that also minimize energy}\label{sec5}

States that give local minima of the output entropy satisfy
equation~\eqref{5}.  To find states that minimize output energy within
each set of connected local minima, we find stationary points of the
Lagrangian
\begin{eqnarray}
&& {\cal L} = ~\tr z a^\dagger a - \mu( -\tr z\ln z - S)
 - \alpha'(\tr \rho -1)\labell{6}\;
\\&&
-\eta'(\tr z -1)
 - \lambda'( - \tr \rho \ln \rho - S_0) 
 - \tr \Lambda' ( z - \Th(\rho) ).
 \nonumber
\end{eqnarray}
Here, $S$ is the output entropy for that set of minima.  A state
within the minimum manifold that minimizes output energy as well as
output entropy must extremize both $\ds$ and ${\cal L}$.

Varying ${\cal L}$ and finding its stationary points shows that such a
state must satisfy the equations
\begin{eqnarray}
&& \mu(\ln z  +1) +  a^\dagger a  - \eta' - \Lambda'= 0
 \nonumber\\&&
 \lambda'(\ln \rho + 1)  - \alpha' +  \Th(\Lambda') =0
 \nonumber\\&&
    \tr \rho = 1,\quad \tr z =1
 \nonumber\\&&
  - \tr \rho \ln \rho = S_0, -\tr z \ln z =S 
 \nonumber\\&&
  z = \Th(\rho) .\labell{7}\;
\end{eqnarray}
To give a local minimum of output entropy, and within that minimum
a local minimum of output energy,
$\rho$ and $z$ must simultaneously solve
equations~\eqref{5} and~\eqref{7} for suitable values of the
Lagrange multipliers.  (We don't have to worry about the minimum
lying on the boundary of the set of density matrices
because points on the boundary have divergent entropy
increase: the set of inputs with minimum output entropy $S$ lies
in the interior of the set of density matrices.)
One solution is obtained by
setting $\mu =0$.  In this case $\Lambda' = a^\dagger a - \eta'$,
and equation~\eqref{7} immediately implies that
$\rho$ and $z$ are thermal states.
All other solutions for $\mu \neq 0$
are also thermal, as will now be shown.

The first two lines of equation~\eqref{5} and the first two
lines of equation~\eqref{7} can now be used to eliminate 
$z$, $\Lambda$, $\Lambda'$ and to obtain an equation for $\rho$:
\begin{eqnarray}
 \hat\lambda (\ln \rho + 1)
-\hat\alpha + \Th(   a^\dagger a/\mu - \hat\eta) = 0.
\labell{8}\;
\end{eqnarray}
Here $\hat\lambda = \lambda'/\mu + \lambda +1$,
$\hat\alpha = \alpha'/\mu+\alpha$, and
$\hat\eta= \eta'/\mu+\eta$. 

Since $\Th(a^\dagger a) = a^\dagger a + \gamma \Delta t$,
equation~\eqref{8} shows that the only possible form for $\rho$ is a
thermal state:
\begin{eqnarray}
\rho = (1 - e^{-\beta})  
e^{-\beta a^\dagger a}.\labell{9}\;
\end{eqnarray}
The output is also thermal:
\begin{eqnarray}
z = (1-e^{-\beta'}) e^{-\beta' a^\dagger a}.
\labell{10}\;
\end{eqnarray}
Here, the inverse temperature $\beta$ is chosen to give the proper
entropy $S_0$ for the input state $\rho$, and $\beta'$ is chosen to
make the energy of the thermal output state equal to the energy for
the thermal input state plus $\Delta E$.  The only minimum energy
stationary point that the Lagrangian technique finds occurs for a
thermal input. We cannot however exclude that other stationary points
exist, that the Lagrangian technique fails to identify.

\section{Thermal states and output entropy}\label{sec6}

For the sake of completeness, we now verify explicitly that this
stationary point corresponds to a minimum of the output entropy $S$.
At the extremum, $\rho$ and $z = \Th(\rho)$ are thermal states.  Look
at perturbations $\Delta \rho, \Delta z$ such that $\rho + \Delta
\rho$ satisfies the input entropy constraint and $z + \Delta z$
satisfies the output energy and dynamical constraints.  The change in
$S$ under these perturbations is
\begin{eqnarray}
&& -\tr \Delta z \ln z + O(\Delta z^2) 
   = \beta \big( \tr (z+\Delta z) a^\dagger a - \tr z a^\dagger a)
 \nonumber\\&&
+ O(\Delta z^2) 
=\beta( E(z+\Delta z) - E(z)) + O(\Delta z^2).\labell{11}\;
\end{eqnarray}
But thermal states not only maximize entropy for fixed energy, they
also minimize energy for fixed entropy.  That is, the energy of
$\rho+\Delta \rho$ is greater than or equal to the energy of $\rho$
for $\Delta \rho \neq 0$.  For the same reason, because $z+\Delta z$
satisfies the output energy constraint, its energy is greater than or
equal to that of $z$.  So the change in $S$ under a non-zero
perturbation of $\rho$ and $z$, equation~\eqref{11}, is non-negative.
(The change is zero for perturbations that respect the symmetry of the
thermal noise.  If our perturbation is a translation in $x$-$p$ space,
we have $\Delta \rho = - i[ \nu a + \bar\nu a^\dagger, \rho]$, and
$\Delta z = - i[ \nu a + \bar\nu a^\dagger, z]$.  In this case $E(z+
\Delta z) = E(z) + O(\Delta z^2)$.)  That is, the stationary point of
$\ds$ corresponds to a local minimum of the output entropy $S$.  

We have shown that thermal inputs give one possible minimum entropy
increase for all cutoffs $N$.  In the limit $N\rightarrow\infty$ (see
the Appendix), thermal states also minimize the rate of entropy
increase, and amongst such states possess a minimum of energy.  For
fixed input entropy, thermal states minimize the output entropy under
infinitesimal thermalization.  Building up non-infinitesimal
thermalization by repeated infinitesimal thermalization, the path that
begins at a thermal state and passes through a sequence of thermal
states is a minimum of output entropy.  Taking the limit that the
input entropy approaches zero (see the Appendix), we find that amongst pure
input states, the zero-energy thermal state or vacuum is also a
minimum for entropy increase under finite thermalization.
\section{Uniqueness of the minimum}
If the gradients of the constraints of the Lagrangian minimizations
are not linearly independent, then not all constrained minima will
satisfy the Lagrange equations. Namely, some minima can be undetected
by this procedure.  Unfortunately, in our second minimization, the
gradients of the constrains are linearly dependent. In fact, in the
first minimization, on the entropy, we are performing a minimization
with the Lagrangian $\Sigma$ of the form
\begin{eqnarray}
\nabla f-\sum_j\lambda_j\nabla g_j=0,\mbox{ with the constraints }g_j=0
\labell{ll}\;,
\end{eqnarray}
where $f$ is the entropy, $g_j$ represents the Lagrange constraints
introduced in Eq.~\eqref{3}, and $\lambda_j$ represents the Lagrange
multipliers introduced in such equation. Then, in the second
minimization, on the energy, we are using the results of the first
minimization with the Lagrangian $\cal L$, i.e. we are performing a
minimization of the form
\begin{eqnarray}
  &&\nabla g-\mu\nabla f-\sum_j\lambda_j'\nabla g_j=0\nonumber\\&&\mbox{
    with the constraints }f=f_{out}, g_j=0 
\labell{lll}\;,
\end{eqnarray}
where $g$ is the energy, $\mu$ and $\lambda'_j$ are the Lagrange
multipliers introduced in Eq.~\eqref{6}, and $f_{out}$ is the minimum
output entropy that results from the first minimization. Replacing
Eq.~\eqref{ll} into~\eqref{lll}, it is then clear that the gradients
of the constraints in the second Lagrange minimization are not
linearly independent, as the gradients of the $g_j$ appear twice
in~\eqref{lll}. Generalized treatments of the Lagrange minimization
theorem (e.g.~see Sec.3.5 of~\cite{hadley}) show that solutions of the
Lagrange equation~\eqref{lll} are all local minima. However, not all
the local minima will solve such equation. This means that, although
thermal states solve this equation, and more specifically, solve
Eq.~\eqref{7}, we are not guaranteed that there might be other
(non-thermal) states that may still be local minima although they do
not solve the Lagrange equations. These minima might have lower output
entropy than the thermal states. There are two ways to conclude the
proof of the conjectures. Either one should prove that thermal states
are the unique constrained minima, or one should prove that any other
constrained minima will have higher output entropy with respect to
thermal states.

\togli{\section{Additivity of minimum output entropy} The result
  immediately extends to multiple modes: in the derivation above, we
  simply take $\rho$ to be an $n$-mode input state, $z$ to be $n$-mode
  output state, and $\Lambda, \Lambda'$ to be $n$-mode Hermitian
  operators.  In the definition of the Lagrangian~\eqref{6}, we
  substitute the $n$-mode Hamiltonian $\sum_{j=1}^n a^\dagger_j a_j$
  for the single mode Hamiltonian $a^\dagger a$.  The derivation now
  goes through as before and yields $n$-mode thermal inputs as the
  states that minimize entropy increase.  That is, the minimum output
  entropy is additive over multiple uses of the channel: entangling
  inputs to the channel cannot reduce it~\cite{10,11,12}.  Not all
  quantum channels have additive minimum output entropy~\cite{11}: the
  methods of our proof show that thermalizing channels of the sort
  discussed here are additive in general.}

\section{Discussion}\label{sec7}
We have introduced a new form of the minimum output entropy conjecture
that refers to infinitesimal channels. We have shown that it is
equivalent to the previous conjecture, although it allows one to
attack it using very different strategies based on variational
techniques. A promising double Lagrangian minimization approach to the
proof of the conjecture has been outlined. Unfortunately, it does not
permit to prove the conjecture because it cannot exclude the presence
of states with output entropy lower than the thermal states that the
double Lagrangian technique identifies.

The proof of the bosonic minimum output entropy
conjecture would establish that the capacity of the bosonic channel
with Gaussian noise and linear loss is attained for coherent state
inputs~\cite{5,6}.  Quantum `tricks' such as squeezing and
entanglement would not enhance the channel's capacity.  Instead,
sending coherent states would be the optimal strategy.  In addition,
the additivity of this channel would be immediately derived. However,
even in this case the optimal detection strategy may well involve
squeezing and entanglement~\cite{5,6}.

The proof of the minimum output entropy conjecture for bosonic
channels in the previous form~\cite{4} or in the form detailed here
remains an important and extremely challenging open question.
Before concluding, it is worth pointing out that some partial progress  have been recently presented in Ref.~\cite{NEW}.

\appendix*\section{}

\subsection{Truncation}

The Lagrangian method requires that the function to be extremized be
continuous on the full input space, and that the gradient of the
function be continuous on the open interior of the input space~\cite{8}.
To insure continuity of entropy and its gradient, we truncate our
input Hilbert space at some high photon number $N$.  Since
infinitesimal thermalization adds at most one photon, the output
Hilbert space is truncated at photon number $N+1$.  To include the
finite truncation in the exposition above, when a Hamiltonian
$a^\dagger a$ appears, this Hamiltonian should be taken to be
$a^\dagger a$ restricted to the appropriate truncated Hilbert space.
That is, when $a^\dagger a$ occurs with an input density matrix
$\rho$, it is the Hamiltonian restricted to the $N$-dimensional input
Hilbert space; when $a^\dagger a$ occurs with an output density matrix
$z$, it is the Hamiltonian restricted to the $N+1$-dimensional output
Hilbert space.

Similarly, the input and output spaces for the operators in our
Lagrangians are defined on the appropriate truncated Hilbert spaces.
The input and output entropies are continuous over the space of
density matrices in the truncated Hilbert spaces, and have continuous
first derivative in the open interior of these spaces.  By making $N$
sufficiently large we insure that any input state with finite energy
is as close as desired to its projection onto the truncated Hilbert
space.  Note that the equations defining the stationary points of the
Lagrangian do not depend on $N$.  That is, for all $N$, the solution
of the Lagrange equations are thermal inputs.  Consequently, in the
limit $N\rightarrow\infty$, this solution must also be thermal.

\togli{Suppose otherwise: then in the infinite dimensional space there is a
non-thermal $\rho_{min}$ such that the entropy increase from
$\rho_{min}$ is strictly less than that for the thermal state
$\rho_{Th}$ with the same input entropy.  By continuity, then, for
sufficiently large $N$, the entropy increase from the projection of
$\rho_{min}$ onto the truncated Hilbert space, $\rho^N_{\min}$, must
be strictly less than that for the truncated thermal state
$\rho_{Th}^N$ with the same entropy.  But this is impossible: we
already proved that $\rho_{th}^N$ gives the minimum output entropy on
the truncated Hilbert space.}

\subsection{The limit as input entropy goes to zero}
We have shown that for all input entropies strictly greater than zero,
$S_0>0$, the path that starts at a thermal state and ends at a thermal
state gives one minimum of the output entropy.  We want to show that
we can extend this path all the way to $S_0=0$.  As the input entropy
goes to zero, the rate of entropy increase diverges.  The only way
that the path that begins at the vacuum (i.e., the zero-temperature
thermal state) could NOT give the minimum increase in entropy would be
if the divergent rate of entropy increase at the boundary contributed
a finite amount $\Delta S$ to the overall increase in entropy as
$\Delta t \rightarrow 0$ in equation~\eqref{2} above, and if $\Delta
S$ were bigger for the vacuum than for some other input state.

We show that this is not the case.  Expanding equation~\eqref{2} to
evaluate $\Delta S$ to next order in $\Delta t$ for input states on
the boundary, we see that the integrated increase in entropy over time
$\Delta t$ goes as $ \Delta S = -\gamma \Delta t \ln \gamma \Delta t$,
which goes to zero in the limit $\Delta t \rightarrow 0$.  In other
words, although the rate of entropy increase diverges at the boundary,
the divergence is a relatively weak, logarithmic one.  In the limit
that $ \Delta t \rightarrow 0$, the entropy increase at the boundary
does not contribute to the overall increase in entropy, even though
the rate of entropy increase diverges.  Accordingly, the path that
starts at the vacuum gives one minimum of the entropy increase.

\noindent{\it Acknowledgements:} This work was supported by the W.M.
Keck foundation, Jeffrey Epstein, NSF, ONR, and DARPA.

\end{document}